**Topological reconstruction of Rubin's multiple imputation via coarse proximity, Seifert–van Kampen gluing and Hurewicz invariants**

Arturo Tozzi (corresponding author)
ASL Napoli 1 Centro, Distretto 27, Naples, Italy
Via Comunale del Principe 13/a 80145
tozziarturo@libero.it

ABSTRACT

Rubin's multiple imputation (MI) generates plausible data completions to account for uncertainty and statistical variability but provides little insight into their global organization. We introduce a topological reconstruction approach that complements MI by examining the structure of the ensemble of completed datasets. Individual imputations are represented as points in a reconstruction space whose coordinates summarize statistical properties of each completed dataset. Then, concepts from coarse geometry and algebraic topology are used to characterize relationships among alternative imputations at multiple scales. Coarse proximity (CP) defines large-scale neighborhoods among admissible reconstructions, generating graphs in which nodes represent completed datasets and edges connect sufficiently similar imputations. Seifert–van Kampen gluing provides conceptual interpretation of how locally related reconstructions assemble into globally coherent structures, whereas Hurewicz-type invariants quantify nontrivial connectivity patterns persisting across reconstruction scales. Synthetic multivariate biomedical datasets representing adult cardiometabolic cohorts were generated with controlled missingness levels. Multiple stochastic imputations were produced for each missingness condition, projected into the reconstruction space and analyzed through CP graphs, connected components, cycle descriptors and scale-dependent topological measures. We found that MI generated structured spaces with distinct connectivity patterns rather than homogeneous clouds of solutions. Topological descriptors were stable despite local numerical variability, whereas increasing missingness produced transitions in reconstruction-space connectivity together with progressive deterioration of reconstruction accuracy. Our approach could be applied to biological networks, neuroscience, systems medicine, ecological modeling, social networks and other domains in which large-scale structural organization contributes to reliable inference.

## INTRODUCTION

Missing data are a pervasive challenge in scientific research, affecting statistical inference, parameter estimation and predictive modelling. Over the past decades, numerous strategies have been developed to address incomplete observations, including case deletion, single imputation methods and likelihood-based procedures (Javanbakht et al. 2022; Raevskiy et al. 2023; Yu et al. 2026). Among these approaches, Rubin's multiple imputation has become one of the most widely adopted solutions because it generates several plausible completions of an incomplete dataset and propagates uncertainty through repeated analyses (Giganti and Shepherd 2020; Yu, Liu, and Peace 2020; Ségalas et al. 2023; Shu et al. 2023). This method has been successfully applied in epidemiology, clinical research, social sciences, economics and many other fields. Despite its broad use, the information extracted from multiple imputation is summarized through conventional statistical descriptors such as means, variances, confidence intervals and prediction errors. These quantities provide limited information regarding the global organization of the collection of admissible completions generated by the imputation process. Indeed, different sets of imputations may exhibit similar statistical summaries while possessing markedly different structural relationships, whereas geometric and topological relationships among these realizations are still largely unexplored. Consequently, these approaches may overlook large-scale organizational properties emerging from interactions among alternative completions.

We introduce a topological reconstruction approach that complements Rubin's multiple imputation by examining the structure of the space formed by admissible completed datasets. We combine concepts from coarse geometry and algebraic topology to characterize relationships among alternative imputations at multiple scales. Individual imputations are represented as points in a reconstruction space whose coordinates summarize relevant statistical properties of the completed datasets. Coarse proximity is used to define large-scale neighborhoods among imputations without relying exclusively on local metric information (Grzegrzolka and Siegert 2023; Shi and Yao 2024). These neighborhoods generate graphs in which nodes correspond to completed datasets and edges connect sufficiently similar reconstructions. The resulting structures are analyzed through topological descriptors inspired by connectedness and cycle formation. In this context, Seifert–van Kampen gluing could provide a conceptual interpretation of how locally related reconstructions assemble into globally coherent structures (Ramos, de Veras, and de Queiroz 2025), while Hurewicz-type invariants could deliver quantitative descriptions of nontrivial connectivity patterns persisting across scales (Arkowitz and Curjel 1964; Sakai 2020; Pilgrim 2025). Rather than focusing exclusively on individual completed datasets, attention is directed toward the organization of the entire ensemble of admissible solutions. To investigate the behavior of our approach, we performed simulations on synthetic multivariate biomedical datasets with controlled missingness levels. Multiple stochastic imputations were generated for each missingness condition and their global organization was examined through reconstruction-space projections, proximity graphs, connectivity measures and scale-dependent topological descriptors. We expected that increasing missingness would alter not only reconstruction accuracy, but also the large-scale structure of admissible completions, potentially producing transitions in the connectivity of the reconstruction space.

We will proceed as follows. First, we describe the simulated datasets, missingness generation procedures, multiple-imputation strategy and topological reconstruction methods. Then, we present the simulation results and quantify the evolution of reconstruction-space topology across missingness levels. Finally, we discuss the implications, limitations and interpretation of our proposed approach.

## METHODS

We studied a simulated cardiometabolic dataset to investigate whether multiple imputation produces not only uncertainty in individual reconstructed values, but also a space of admissible data completions that can be characterized through topological descriptors. We used stochastic regression-based multiple imputation, dimensional reduction, coarse proximity graphs, connected-component analysis, cycle counts, Seifert–van Kampen-inspired gluing, Hurewicz-type descriptors and scale-dependent topological summaries. We aimed to identify an experimentally discriminable signature: a transition in the topology of the imputation space as missingness increased, expressed by changes in connected components, independent cycles, reconstruction error and classification variability.

**Synthetic cohort**. A virtual cohort of 240 subjects was generated to mimic a compact biomedical dataset with continuous clinical variables representative of adult cardiometabolic cohorts. Each subject was assigned age (years), body mass index (kg/m$^2$), fasting glucose (mg/dL), systolic blood pressure (mmHg), low-density lipoprotein cholesterol (mg/dL), high-density lipoprotein cholesterol (mg/dL) and C-reactive protein (mg/L) (Yu et al. 2018; Chung 2019; Çubukçu and Topcu 2022; Oh et al. 2022; Qiao et al. 2023; Mitani et al. 2023; Solianik et al. 2023; Tsang et al. 2023; Chin, Potter, and Friedl 2024). Variable ranges and dependence structures were chosen to be consistent with reported distributions of established cardiometabolic risk factors in population-based studies (D'Agostino et al. 2008; Marino et al. 2014). The simulated observations were generated from correlated latent cardiometabolic risk variables to reproduce the well-documented associations among adiposity, systemic inflammation, glycemic status, blood pressure and lipid metabolism.
For each virtual subject $i$, a latent risk variable $r_i$ was defined as a weighted combination of age, body mass index and C-reactive protein, with an added Gaussian error term:

$$r_i = 0.04(a_i - 55) + 0.12(b_i - 27) + 0.08(c_i - 3) + \varepsilon_i,$$

where $a_i$ denotes age, $b_i$ body mass index, $c_i$ C-reactive protein and $\varepsilon_i$ random variability. Clinical variables were then generated as functions of $r_i$ plus independent noise. For example, glucose was generated as:

$$g_i = 92 + 10r_i + 0.35(b_i - 27) + \eta_i,$$

and systolic blood pressure as:

$$s_i = 118 + 8r_i + 0.45(a_i - 55) + \xi_i.$$

All simulated variables were clipped to clinically plausible ranges.

**Missingness design**. Controlled missingness was introduced at six levels: 5%, 10%, 20%, 30%, 40% and 50%. Missingness was not imposed as purely random deletion. Instead, glucose, systolic blood pressure and low-density lipoprotein cholesterol were made more likely to be missing in subjects with higher latent risk. This choice generated a missing-at-random-like structure in which the probability of observing a missing value depended on other measurable subject properties. If $m_{ij}$ denotes the missingness indicator for subject $i$and variable $j$, the probability of missingness was modeled as:

$$P(m_{ij} = 1) = \rho(0.35+1.3q_i),$$

where $\rho$ is the nominal missingness level and $q_i$ is the normalized latent risk score. For variables less directly associated with cardiometabolic risk, lower missingness probabilities were used. This procedure produced incomplete matrices:

$$X^{(\rho)} = \{x_{ij}^{(\rho)}\},$$

where some entries were unobserved. The complete matrix $X$ was retained only for simulation validation, allowing reconstruction error to be calculated after imputation. In turn, the incomplete matrix was the object analyzed by the imputation procedure.

**Multiple imputation**. For each missingness level, forty imputed datasets were generated. This ensemble size was chosen to provide stable estimates of between-imputation variability while yielding a sufficiently dense sampling of the space of admissible data completions for the computation of our topological descriptors. Missing entries were first replaced by median values to obtain an initial complete matritx, then each variable with missing values was imputed by stochastic regression using the remaining variables as predictors. For variable $j$, observed and missing rows were separated. A regression model was fitted on observed entries:

$$x_{ij} = \alpha_j + \sum_{k \neq j} \beta_{jk} x_{ik} + e_{ij}.$$

For rows with missing $x_{ij}$, the conditional mean was estimated as:

$$\hat{x}_{ij} = \hat{\alpha}_j + \sum_{k \neq j} \hat{\beta}_{jk} x_{ik}.$$

To avoid producing identical deterministic completions, a stochastic residual term was added:

$$x_{ij}^* = \hat{x}_{ij} + z_{ij}\hat{\sigma}_{ij},$$

where $z_{ij}$ was sampled from a standard normal distribution and $\hat{\sigma}_{ij}$ represented predictive uncertainty. This procedure yielded forty imputed datasets for each missingness level:

$$\mathcal{I}_\rho = \{X_\rho^{(1)}, X_\rho^{(2)}, \dots, X_\rho^{(40)}\}.$$

Each completed dataset retained the same units as the original variables. Values were constrained to remain within plausible clinical intervals.

**Reconstruction space**. Each completed dataset was summarized by clinically interpretable statistics. For imputation $k$ at missingness level $\rho$, we computed the mean fasting glucose, mean systolic blood pressure, mean low-density lipoprotein cholesterol, mean body mass index and estimated metabolic-syndrome prevalence. This produced a summary vector:

$$v_{\rho k} = (\bar{g}_{\rho k}, \bar{s}_{\rho k}, \bar{\ell}_{\rho k}, \bar{b}_{\rho k}, p_{\rho k}).$$

The set of vectors formed a reconstruction space of admissible completions. Standardization was applied to remove unit dominance:

$$z_{\rho kj} = \frac{v_{\rho kj} - \mu_j}{\sigma_j}.$$

Principal component analysis was then used only as a visualization and projection method. Given the standardized matrix $Z$, the covariance matrix was computed as:

$$C = \frac{1}{n-1} Z^T Z.$$

Eigenvectors of $C$ defined orthogonal reconstruction axes. Each imputation was projected into the first two axes:

$$y_{\rho k} = Z_{\rho k} W_2.$$

The resulting two-dimensional coordinates represented each completed dataset as one point in reconstruction space.

**Coarse proximity**. Coarse proximity was used to define large-scale similarity among imputed datasets (Grzegrzolka and Siegert 2019, 2023). Instead of requiring exact local equality, two completions were considered related when their projected clinical profiles were sufficiently close at a chosen scale. For imputations $k$ and $l$, Euclidean distance in standardized reconstruction space was calculated as:

$$d_\rho(k,l) = \sqrt{(y_{\rho k1} - y_{\rho l1})^2 + (y_{\rho k2} - y_{\rho l2})^2}.$$

A coarse proximity relation was defined by a radius $\epsilon$:

$$k \sim_\epsilon l \text{ if and only if } d_\rho(k,l) \leq \epsilon.$$

This relation generated a graph:

$$G_{\rho,\epsilon} = (V_\rho, E_{\rho,\epsilon}),$$

where vertices were imputed datasets and edges joined pairs of sufficiently similar completions. Coarse proximity was used because multiple imputations can differ in small local numerical details while still belonging to the same broad reconstruction region. Therefore, the proximity graph represented the large-scale organization of admissible completions rather than individual variable-level differences alone.

**Connectedness**. Connected components were computed from the adjacency matrix of the coarse proximity graph (Prolubnikov 2023). The adjacency matrix was defined as:

$$A_{kl} = \begin{cases} 1, & d_\rho(k,l) \le \epsilon, \\ 0, & d_\rho(k,l) > \epsilon. \end{cases}$$

The number of connected components, denoted $\beta_0$, quantified fragmentation of the imputation space. A high $\beta_0$ indicated that admissible completions separated into multiple disconnected families. A low $\beta_0$ indicated that completions formed a more coherent reconstruction region. The first cycle count was calculated through the graph-theoretic relation:

$$\beta_1 = |E| - |V| + \beta_0,$$

where $|E|$ is the number of edges and $|V|$ the number of vertices. This quantity measures independent cycles in the proximity graph. In the present setting, a cycle indicates that several imputations are mutually connected through alternative paths, producing a non-tree-like organization of the completion space. Both $\beta_0$ and $\beta_1$ were evaluated across missingness levels and across proximity radii. This allowed us to distinguish numerical dispersion from topological reorganization.

**Gluing principle**. Seifert–van Kampen theory was used as a guiding topological principle for interpreting how local reconstruction regions combine into a global imputation space. In algebraic topology, the theorem describes how the fundamental group of a space can be obtained from two overlapping subspaces when their intersection is sufficiently connected. In simplified terms, if a space $X$ is covered by two regions $U$ and $V$, then global loop structure can be reconstructed from the loop structures of $U$, $V$ and $U \cap V$. Symbolically, the classical idea is expressed as:

$$\pi_1(X) = \pi_1(U) * \pi_1(V) \; / \; \langle \text{relations from } U \cap V \rangle.$$

In our simulation, local groups of similar imputations were interpreted as patches. Edges among patches represented overlaps or coarse compatibility. We did not perform an explicit computation of fundamental groups. Instead, we used the gluing principle of the Seifert–van Kampen theorem operationally, interpreting the progressive merging of previously disconnected imputation neighborhoods into larger connected structures as the coarse-proximity radius increased. The decreasing component count measured this gluing process. The aim was to observe whether local imputational neighborhoods assembled gradually or through abrupt transitions.

**Hurewicz descriptor**. The Hurewicz theorem links homotopy information, which concerns continuous deformation of paths and loops, with homology information, which counts holes and cycles through algebraic invariants. In its classical form, under suitable connectivity assumptions, the first nontrivial homotopy group maps to the corresponding homology group. The conceptual bridge is:

$$\pi_n(X) \longrightarrow H_n(X).$$

In the simulation, the full homotopy structure of the imputation space was not computed. Instead, a Hurewicz-like descriptor was defined by tracking graph cycles as homological signatures of nontrivial connectivity. The first cycle count $\beta_1$ was used as an accessible representation of loop-like organization among admissible completions. This choice was appropriate for a finite set of imputed datasets represented as a proximity graph. A tree-like graph has $\beta_1 = 0$, whereas additional compatible paths among imputations increase $\beta_1$. Therefore, changes in $\beta_1$ across scales described whether the reconstruction space remained simple, fragmented or loop-rich. The descriptor was not interpreted as a biological invariant, but as a mathematical summary of the imputation ensemble.

**Classification metric**. To relate reconstruction topology to an interpretable simulated clinical outcome, each completed dataset was used to estimate metabolic-syndrome-like prevalence. A subject was classified as positive when at least two of the following simulated criteria were met: body mass index of at least 30 kg/m², fasting glucose of at least 126 mg/dL, systolic blood pressure of at least 130 mmHg, low-density lipoprotein cholesterol of at least 160 mg/dL or high-density lipoprotein cholesterol below 40 mg/dL. For subject $i$, the classification score was:

$$c_i = \mathbf{1}(b_i \ge 30) + \mathbf{1}(g_i \ge 126) + \mathbf{1}(s_i \ge 130) + \mathbf{1}(\ell_i \ge 160) + \mathbf{1}(h_i < 40).$$

The binary classification was:

$$q_i = \begin{cases} 1, & c_i \ge 2, \\ 0, & c_i < 2. \end{cases}$$

For each imputation $k$, prevalence was:

$$p_{\rho k} = 100 \times \frac{1}{N}\sum_{i=1}^{N} q_{\rho k i}\,.$$

The standard deviation of $p_{\rho k}$ across forty imputations quantified classification variability induced by missing data.

Error estimation. Because our simulation retained the original complete dataset, reconstruction accuracy could be calculated directly. For each missingness level, the average imputed value for each subject and variable was computed across forty completions. For glucose, the root mean square error was:

$$\mathrm{RMSE}_{g} = \sqrt{\frac{1}{N}\sum_{i=1}^{N}(\hat{g}_i - g_i)^2}.$$

For systolic blood pressure, the corresponding expression was:

$$\mathrm{RMSE}_{s} = \sqrt{\frac{1}{N}\sum_{i=1}^{N}(\hat{s}_i - s_i)^2}.$$

These errors were reported in mg/dL and mmHg. The error analysis was kept separate from the topological analysis. Reconstruction error measured numerical deviation from the known simulated truth, whereas $\beta_0$, $\beta_1$ and proximity graphs measured the shape of the ensemble of admissible completions. This separation allowed our simulation to compare conventional accuracy summaries with topological summaries of imputation structure.

**Software tools**. All simulations, imputations, numerical analyses and figures were generated in Python. NumPy was used for numerical array operations, random sampling and mathematical transformations. Pandas was used to organize the simulated clinical dataset and export the summary table. Scikit-learn was used for median initialization, Bayesian ridge regression and principal component analysis. SciPy was used to compute pairwise distances and connected components of proximity graphs.

## RESULTS

We report here the behavior of stochastic multiple imputation and topological reconstruction across progressively increasing levels of missing data. Our analyses focused on the geometry of admissible completions, coarse-proximity connectivity, cycle formation, reconstruction accuracy and variability of a simulated clinical classification. Quantitative results are presented for missingness levels ranging from 5% to 50%, with forty completed datasets generated for each condition.

**Reconstruction topology**. Throughout the investigated missingness range, the space of admissible imputations exhibited an organized rather than random structure. Projection of the forty imputations generated at each missingness level into the reconstruction space revealed partially overlapping but distinguishable clusters corresponding to different percentages of missing data (Figure 1, upper-left inset). The number of connected components varied between 3 and 9 across the investigated missingness levels (Figure 1, upper-right inset). Six connected components were observed at 5% and 10% missingness, followed by seven at 20%, three at 30%, nine at 40% and five at 50%, indicating a non-monotonic evolution of the connectivity of the reconstruction space. The first cycle count varied between 56 and 76 across the investigated missingness levels, reaching 56 at 5% missingness and a maximum of 76 at 40% missingness. A Spearman correlation analysis between missingness percentage and cycle count did not reveal a significant monotonic trend ($\rho = 0.314$, $p = 0.544$), indicating that topological complexity evolved nonlinearly rather than increasing proportionally with missingness. The dataset displayed positive associations among body mass index, glucose, systolic blood pressure and low-density lipoprotein cholesterol, whereas high-density lipoprotein cholesterol showed negative correlations with most cardiometabolic variables (Figure 1, lower-right inset). Examination of the coarse-proximity graph at 40% missingness revealed multiple densely connected local neighborhoods linked by a smaller number of bridging edges (Figure 2, left inset). The coexistence of connected clusters and numerous cycles indicates that admissible completions formed structured families rather than isolated realizations, suggesting that reconstruction uncertainty displayed a measurable topological organization beyond simple numerical dispersion.

**Accuracy and transitions**. Reconstruction error increased progressively as larger fractions of data were removed (Figure 1, lower-left inset). Mean glucose root mean square error rose from 2.11 mg/dL at 5% missingness to 8.66 mg/dL at 50% missingness. Mean systolic blood pressure root mean square error increased from 1.94 mmHg to 9.03 mmHg across the same interval. Spearman correlation analysis demonstrated a perfect monotonic relationship between missingness and reconstruction error for both glucose ($\rho = 1.000$, $p < 0.001$) and systolic blood pressure ($\rho = 1.000$, $p < 0.001$). Variability of estimated metabolic-syndrome prevalence also increased with missingness (Figure 1, lower-middle inset). The standard deviation across imputations rose from 0.54% at 5% missingness to 1.54% at 50% missingness. A significant positive Spearman association was observed between missingness and classification variability ($\rho = 0.829$, $p = 0.042$). Scale-dependent analysis of the proximity graph provided additional information regarding the organization of admissible completions (Figure 2, upper-right inset). As the coarse-proximity radius increased from 0.25 to 1.80, the number of connected components decreased from 31 to a single connected structure, indicating progressive gluing of initially separated reconstruction regions. Simultaneously, the number of independent cycles increased from 5 to 264 (Figure 2, lower-right inset), showing the emergence of increasingly rich connectivity patterns as additional links were established among imputations. The transition from numerous disconnected groups to a single connected object occurred over a relatively narrow interval of proximity radii, indicating that the global organization of admissible completions depended strongly on the observational scale used to define similarity relationships. Together, the numerical and topological analyses showed that increasing missingness altered both reconstruction fidelity and the large-scale structure of the imputation ensemble.

Overall, our results show that multiple imputation gives rise to structured spaces of admissible data completions whose organization can be quantified independently of conventional reconstruction accuracy measures. Reconstruction error and classification variability increased with missingness, whereas connected components and cycle counts evolved nonlinearly. Scale-dependent coarse-proximity analysis revealed transitions from fragmented local neighborhoods to globally connected reconstruction spaces. These findings indicate that conventional statistical summaries can be complemented by descriptors of large-scale connectivity and topological organization.

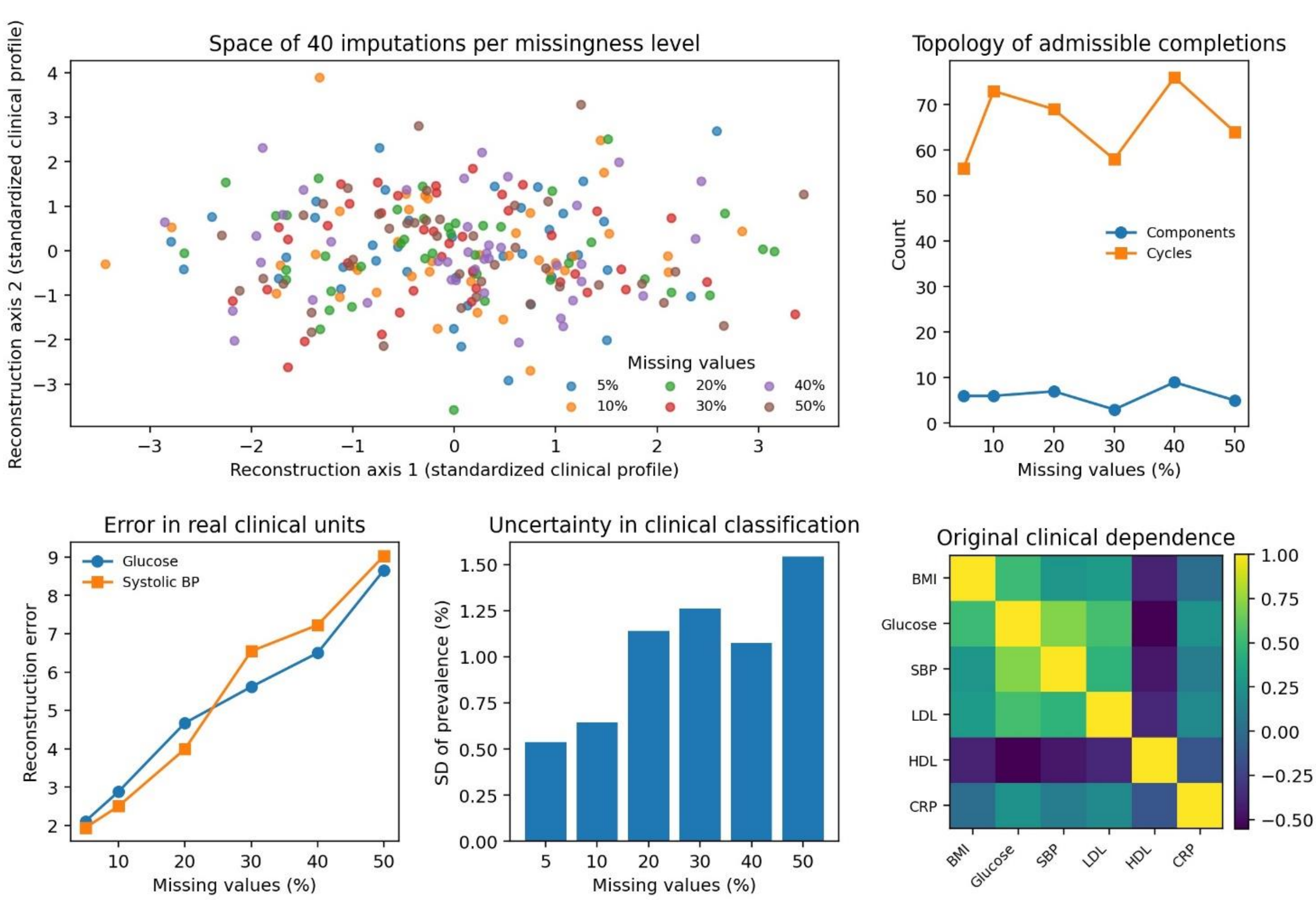

**Figure 1**.Topological multiple imputation in a simulated cardiometabolic cohort under increasing missingness. The figure summarizes the behavior of our topological reconstruction approach in a simulated cohort of 240 virtual subjects characterized by age (years), body mass index (kg/m²), fasting glucose (mg/dL), systolic blood pressure (mmHg), low-density lipoprotein cholesterol (mg/dL), high-density lipoprotein cholesterol (mg/dL) and C-reactive protein concentration (mg/L). Missing values were introduced at rates ranging from 5% to 50% and forty stochastic imputations were generated for each missingness level, yielding ensembles of admissible completed datasets whose large-scale organization was subsequently analyzed through topological descriptors.

**Upper-left inset**: reconstruction space of all admissible imputations after dimensionality reduction. Each point corresponds to one completed dataset and is positioned according to its overall clinical profile. Colors indicate the proportion of missing values. The distribution of points provides a visual representation of the geometry of the imputation space and the degree of separation among alternative reconstructions.
**Upper-right inset**: topological descriptors derived from coarse-proximity relationships among admissible imputations. The number of connected components reflects the fragmentation of the reconstruction space, whereas the number of independent cycles represents Hurewicz-like topological invariants characterizing nontrivial connectivity patterns.
**Lower-left inset**: reconstruction error expressed in clinical units. Curves correspond to the root mean square reconstruction error of fasting glucose (mg/dL) and systolic blood pressure (mmHg) relative to the complete dataset. Both variables exhibited a progressive increase in reconstruction error with increasing missingness, indicating a monotonic deterioration in numerical reconstruction accuracy
**Lower-middle inset**: Standard deviation of estimated metabolic-syndrome prevalence across the forty imputations, providing a measure of classification uncertainty generated by missing data. The standard deviation increased from 0.54% at 5% missingness to 1.54% at 50% missingness, indicating progressively greater variability in prevalence estimates as the proportion of missing observations increased.
**Lower-right inset**: Pearson correlation matrix of the original complete dataset, illustrating the dependence structure among the simulated clinical variables. Positive correlations are observed among body mass index, fasting glucose, systolic blood pressure, low-density lipoprotein cholesterol and C-reactive protein, whereas high-density lipoprotein cholesterol exhibits predominantly negative correlations with the remaining variables. Pairwise correlation coefficients ranged from approximately **−0.45 to +0.62**, indicating a moderately correlated multivariate structure that served as the statistical basis for stochastic imputation and influenced the geometry of the admissible reconstruction space.

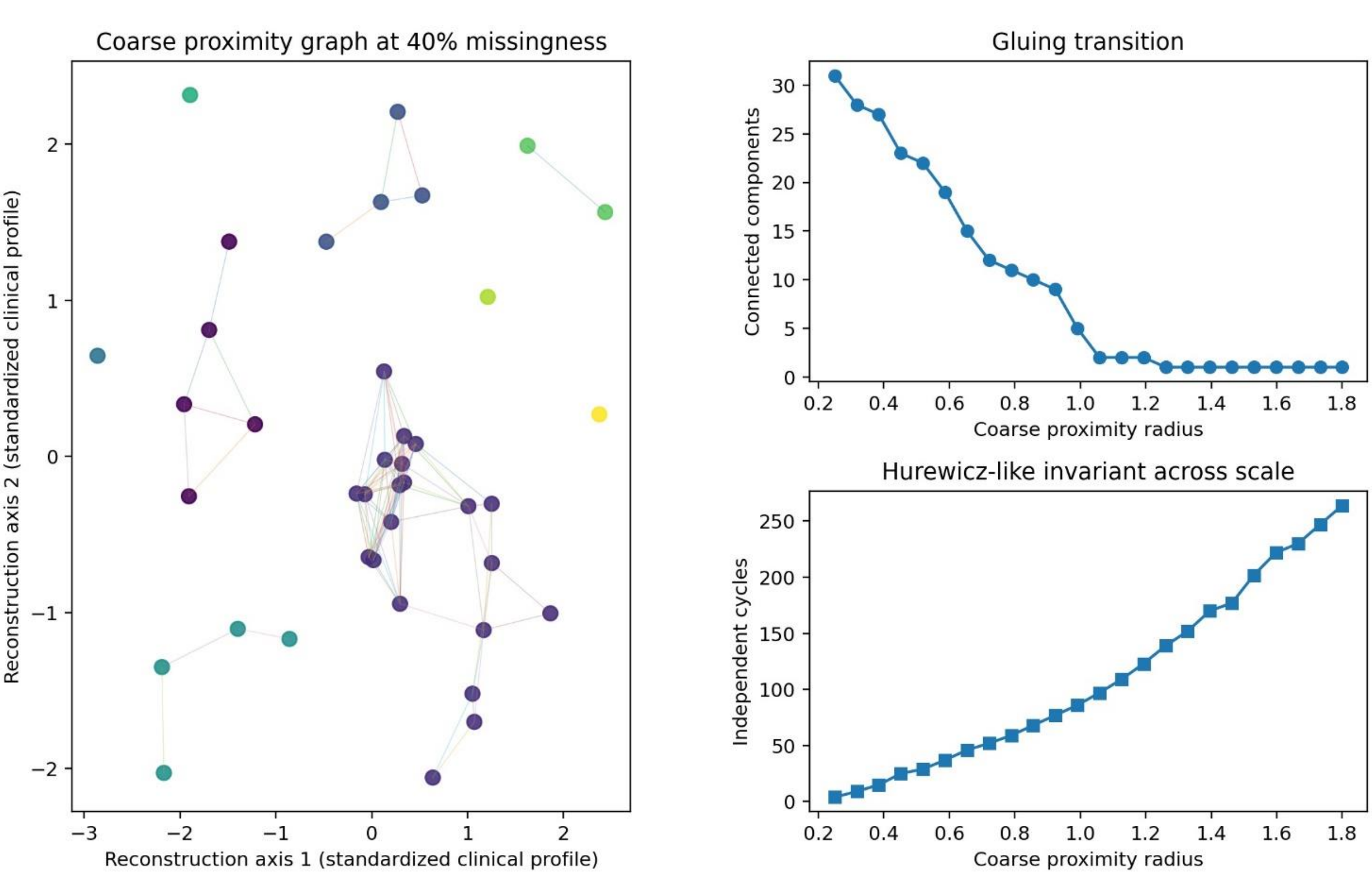

**Figure 2.** Scale-dependent topology of admissible imputations and coarse-proximity reconstruction. The figure illustrates how the global organization of admissible imputations evolves as the scale of coarse proximity is varied. Results are shown for the scenario with 40% missing values.
**Large left inset**: coarse-proximity graph constructed from forty completed datasets projected into the reconstruction space. Each node represents an individual imputed dataset and edges connect pairs of imputations separated by less than a predefined proximity radius. Node positions correspond to standardized clinical reconstruction coordinates. Clusters identify groups of mutually similar imputations, whereas isolated nodes represent alternative reconstructions that remain weakly connected to the dominant solution families.

**Upper-right inset**: gluing transition. As the coarse-proximity radius increases, initially disconnected groups of admissible imputations progressively merge into larger connected structures. The decreasing number of connected components quantifies the transition from fragmented local reconstructions toward a globally integrated reconstruction space, analogous to the assembly of local patches into a coherent global object.
**Lower-right inset**: evolution of the Hurewicz-like invariant across increasing proximity scales. The number of independent cycles increases as additional connections are established among imputations, revealing the emergence of nontrivial topological organization within the reconstruction space.

## CONCLUSIONS

We asked whether collections of datasets generated by Rubin's multiple imputation contain an internal large-scale organization independent of conventional statistical uncertainty measures. Specifically, we investigated whether admissible completions form structured spaces whose properties can be described through coarse proximity, connectedness and topological invariants. Our simulations compared multiple missingness levels in a synthetic cardiometabolic cohort while preserving known ground truth values, allowing simultaneous assessment of reconstruction accuracy and topological organization. We found that increasing missingness modified both the numerical reconstruction performance and the geometry of the admissible completion space. Reconstruction errors increased progressively with missingness, while the organization of imputations exhibited changes in connected components, cycle counts and scale-dependent connectivity patterns. The resulting spaces were neither fully fragmented nor completely homogeneous, but displayed structured families of admissible reconstructions connected through identifiable proximity relationships. Therefore, uncertainty associated with missing data could be examined not only through distributions of estimated values, but also through relationships among alternative completed datasets. Our results suggest that topological descriptors and conventional statistical summaries capture different aspects of the reconstruction process: whereas reconstruction error quantifies deviation from known values, topological descriptors characterize how alternative completions are arranged relative to one another.

Our approach differs from existing missing-data methodologies in several respects. Techniques such as listwise deletion, mean substitution, expectation-maximization algorithms, inverse probability weighting and standard multiple imputation focus primarily on parameter estimation, prediction or uncertainty quantification (Lieberman-Betz et al. 2014; Chan 2017; De Raadt et al. 2019; Chen et al. 2024; Niloofar, Aghdam, and Eslahchi 2024). In contrast, our approach analyzes relationships among completed datasets themselves. Unlike principal component analysis, cluster analysis and conventional distance-based visualization methods, which primarily describe the geometric distribution of observations (Martins et al. 2019; Mi et al. 2019; Ben Salem and Ben Abdelaziz 2021; Jang 2021), our procedure quantifies the topology of the imputation space through measures of connectedness and independent cycle structure. Compared with Bayesian posterior summaries, which describe probability distributions of parameters (Alquier 2020; Huber, Georgia, and Finley 2023; van Dijk et al. 2023; Golchi and Willard 2024), our approach investigates the organization of admissible reconstructions generated by those distributions. Compared with sensitivity analyses based on repeated imputations (Crameri et al. 2015; Mehrotra, Liu, and Permutt 2017; Tompsett et al. 2020), our descriptors are capable of distinguishing situations with comparable numerical variability but different global arrangements of solutions. The use of coarse proximity allows examination of large-scale relationships without requiring strict local similarity, while Seifert–van Kampen-inspired gluing and Hurewicz-type descriptors provide quantitative summaries of connectivity across scales (Bhardwaj and Tyagi 2019; Christensen and Scoccola 2023).

Our study has limitations. First, the analyses were conducted entirely on simulated data. Second, the simulated cohort contained a restricted set of continuous variables with predefined correlation structures and did not include categorical variables, longitudinal observations or multimodal measurements. Third, the missingness mechanism was designed to resemble clinically plausible conditions but cannot represent the full diversity of real-world missing-data processes. Fourth, the topological descriptors were derived from finite graphs constructed in reduced-dimensional reconstruction spaces; alternative distance measures, dimensional reductions or graph-construction procedures could produce different quantitative values. Fifth, the Seifert–van Kampen and Hurewicz concepts were used as operational tools for describing connectivity and cycle structure rather than as complete algebraic-topological calculations. Sources of potential bias include the selected proximity thresholds, the stochastic characteristics of the imputation algorithm and the peculiar summary statistics used to define reconstruction coordinates. Unresolved questions concern the stability of topological descriptors under different imputation models, the behavior of the method in very high-dimensional datasets, the relationship between topological signatures and missingness mechanisms, and the extent to which different biomedical domains produce comparable reconstruction-space organizations.

Our approach suggests experimentally testable previsions. First, empirical datasets with increasing fractions of artificially removed observations should exhibit a monotonic increase in reconstruction error while simultaneously displaying non-monotonic variation in topological descriptors. Specifically, the relationship between missingness and cycle counts is expected to be weaker than the relationship between missingness and reconstruction error.

Second, biological datasets generated from strongly constrained systems should produce fewer connected components than datasets generated from more heterogeneous populations when analyzed at identical missingness levels and proximity scales.
Third, repeated imputations generated by different algorithms may produce similar average estimates while exhibiting distinct connectivity structures. This prediction can be tested by comparing graph-derived descriptors obtained from methods such as predictive mean matching, Bayesian regression, random forests and neural-network imputations.
Fourth, increasing the coarse-proximity radius should induce a reproducible transition from multiple disconnected reconstruction regions toward a single connected structure. The critical radius associated with this transition constitutes a measurable observable that can be compared across datasets.
Fifth, if reconstruction spaces contain persistent organizational features, the values of connected-component counts and cycle counts should remain more stable under repeated resampling than under randomized permutation of imputation labels.
Sixth, datasets characterized by stronger internal correlation structures should exhibit lower critical gluing radii and denser proximity graphs than datasets with weaker dependence structures.
Future research may investigate persistent homology, higher-dimensional topological invariants, alternative notions of coarse similarity, information-theoretic descriptors, longitudinal missing-data processes, network-valued observations and comparisons between synthetic and empirical datasets.
Our analysis is applied as a post-processing step to completed datasets generated by conventional multiple-imputation procedures and therefore does not require modifications to existing data-processing pipeline. Reconstruction coordinates can be derived from summary statistics already reported in many studies, allowing direct computation of proximity graphs and associated descriptors. Our method can be integrated with existing quality-control procedures to compare different imputation strategies, evaluate reconstruction consistency across repeated analyses and identify transitions in reconstruction-space organization.

In conclusion, we investigated missing-data reconstruction from the perspective of the organization of admissible completed datasets. Multiple imputations generated structured spaces whose connectivity and cycle properties were quantified independently of conventional measures of reconstruction accuracy. We showed that numerical uncertainty and topological organization could provide distinct descriptions of the same reconstruction process. Therfore, examination of admissible completions as interconnected objects could extend the characterization of missing-data solutions beyond parameter estimates and error metrics alone.

DECLARATIONS

**Ethics approval and consent to participate.** Ethics approval statement is not applicable. This research does not contain any studies with human participants or animals performed by the Author.
**Consent for publication.** The Author transfers all copyright ownership, in the event the work is published. The undersigned author warrants that the article is original, does not infringe on any copyright or other proprietary right of any third part, is not under consideration by another journal and has not been previously published.
**Availability of data and materials.** All data and materials generated or analyzed during this study are included in the manuscript. The Author had full access to all the data in the study and took responsibility for the integrity of the data and the accuracy of the data analysis.
**Disclaimer**. The views expressed are those of the author and do not necessarily reflect those of the affiliated institutions.
**Competing interests.** The Author does not have any known or potential conflict of interest including any financial, personal or other relationships with other people or organizations within three years of beginning the submitted work that could inappropriately influence or be perceived to influence their work.
**Funding.** This research did not receive any specific grant from funding agencies in the public, commercial or not-for-profit sectors.
**Acknowledgements:** none.
**Authors' contributions.** The Author performed: study concept and design, acquisition of data, analysis and interpretation of data, drafting of the manuscript, critical revision of the manuscript for important intellectual content, statistical analysis, obtained funding, administrative, technical and material support, study supervision.
**Declaration of generative AI and AI-assisted technologies in the writing process.** During the preparation of this work, the author used ChatGPT 5.3 to assist with data analysis and manuscript drafting and to improve spelling, grammar and general editing. After using this tool, the author reviewed and edited the content as needed, taking full responsibility for the content of the publication.